\documentclass[10pt]{article}

\usepackage[margin=1in]{geometry}

\usepackage{graphicx}
\usepackage{dcolumn}
\usepackage{bm}
\usepackage{hyperref}
\usepackage[mathlines,pagewise]{lineno}

\usepackage{mathptmx}

\usepackage{titlesec}
\titleformat{\section}{\large\bfseries}{\thesection.}{1em}{}[]
\titleformat{\subsection}{\normalsize\itshape}{\thesubsection.}{1em}{}[]
\titlespacing{\section}{0pt}{7pt}{0pt}[0pt]
\titlespacing{\subsection}{0pt}{3pt}{0pt}[0pt] 
\usepackage[font=footnotesize,labelfont=bf]{caption} 

\usepackage[numbers,square,sort&compress,comma]{natbib} 

\usepackage[color=blue!20]{todonotes}
\usepackage{xcolor}

\begin{document}

\begin{center}

\Large \textbf{Coordinated tractions {increase} the size of a collectively moving pack in a cell monolayer}

\normalsize
Aashrith Saraswathibhatla,$^1$
Silke Henkes,$^2$
Emmett E. Galles,$^1$
Rastko Sknepnek,$^{3,4}$
Jacob Notbohm$^1$

$^1$Department of Engineering Physics, University of Wisconsin-Madison, Madison, Wisconsin 53706, USA\\
$^2$School of Mathematics, University of Bristol, Bristol BS8 1TW, United Kingdom\\
$^3$School of Science and Engineering, University of Dundee, Dundee DD1 4HN, United Kingdom\\
$^4$School of Life Sciences, University of Dundee, Dundee DD1 5EH, United Kingdom

\end{center}

Correspondence should be sent to: jknotbohm@wisc.edu

\section*{Keywords}

Collective cell migration, traction persistence, traction alignment, Self-propelled Voronoi model

\section*{Abstract}
Cells in an epithelial monolayer coordinate motion with their neighbors giving rise to collectively moving packs of sizes spanning multiple cell diameters. The physical mechanism controlling the pack size, however, remains unclear. A potential mechanism comes from assuming that cell--substrate traction forces persist over some time scale: with large enough persistence time, collective cell packs emerge. To test this {hypothesis}, we measured the velocity and net traction of each cell. The data showed that in addition to having some temporal persistence, tractions were spatially correlated, suggesting that cells coordinate with their neighbors to apply tractions in the same direction. Chemical inhibitors and activators of actomyosin contraction were used to determine effects of altering the traction persistence and alignment. {Numerical simulations based on the self-propelled Voronoi model, augmented to include both traction persistence and alignment and calibrated against the experimental data, matched the experimentally measured pack size. The model identified that if there were no alignment of traction between neighboring cells, the size of the collective pack would be substantially smaller than observed in the experiments. Hence, combining experiments and a simple mechanical model,} this study confirms the long-standing assumption of traction persistence and adds the notion of traction alignment between neighbors. Together, persistence and alignment {are two factors controlling} the size of a collectively moving cell pack.

\setcounter{section}{0}
\section{Introduction}
During tissue formation and repair, cells migrate in collective groups and packs \cite{friedl2009collective}. The characteristic size of a collective cell pack lies at an intermediate point between extremes present in other collective systems---cell motion is not fully random as it would be in systems driven by thermal fluctuations \cite{liu1998jamming}, nor is the motion highly coordinated as in flocks of birds \cite{ballerini2008interaction} or schools of fish \cite{ward2008quorum}. 
Instead, in collective cell swirls, packs, and waves, the motion is correlated over length scales ranging from a few cells to a couple dozen cells \cite{poujade2007collective, angelini2011glass, serra2012mechanical, deforet2014emergence, das2015molecular, garcia2015physics,  notbohm2016cellular, loza2016cell, malinverno2017endocytic, petrolli2019confinement, henkes2020dense, hino2020erk}. Hence, there must be a mechanism for correlation of velocity between neighboring cells.

A plausible scenario, {from predictions of prior models of epithelial cell sheets}, is that the finite time required for a cell to reorganize its force-generating machinery leads to a temporal persistence to each cell's motion. The models, including both geometric (i.e., vertex) models \cite{farhadifar2007influence, bi2016motility, barton2017active, mitchel2020primary} and simpler particle-based models \cite{Szabo2006, Drasdo2007, henkes2020dense}, have shown that if the persistence time is sufficiently large, collective packs can emerge \cite{bi2016motility, henkes2020dense, mitchel2020primary}. While appealing for its simplicity, this explanation includes no explicit mechanism for a cell to polarize and align its propulsive force with that of its neighbors.
Cell-induced substrate displacements are correlated over multiple cell lengths \cite{angelini2010cell}, implying that there may exist some mechanism of cell coordination. Several physical mechanisms have been proposed, including a tendency of cells to migrate along the local orientation of the maximal principal stress \cite{tambe2011collective}, a tendency to apply tractions that pull each cell toward regions of empty space \cite{kim2013propulsion}, {force transmission across the cell layer that causes collective stiffness sensing \cite{sunyer2016collective}, collective cell migration in response to electric fields \cite{cohen2014galvanotactic}, and alignment between neighboring cells creating} orientational nematic order \cite{saw2017topological, kawaguchi2017topological, duclos2017topological}.
Models of such coordination have suggested that a coupling between the force on a cell and its traction results in flock formation \cite{Szabo2006, giavazzi2018flocking}.
In addition, cells can coordinate to align stress fibers in the same direction \cite{nava2020heterochromatin, lopez2020apical, popkova2020cdc42}, and there exist molecular mechanisms that enable neighboring cells to coordinate front-back polarization \cite{das2015molecular, hino2020erk}.
These observations give circumstantial evidence that cells may coordinate with their neighbors to apply tractions in the same direction. Coordination of traction would then be a second mechanism—in addition to temporal persistence—that controls the size of a collective cell pack. Direct evidence, however, is lacking, because neither the temporal persistence time nor a coordination of traction forces between neighbors has been quantified experimentally at the single cell level.

In this letter, using traction force microscopy and cell tracking, we developed a method to measure the net traction exerted by individual cells over time. Using this new single-cell measure, we verified the long-standing notion that cell tractions are persistent over time. Importantly, a spatial autocorrelation of single cell traction demonstrated that the collective motion is not controlled merely by a temporal persistence of traction; rather, cells also coordinate to align directions of traction with those of their neighbors. Traction persistence and alignment were investigated further, by using chemical perturbations of actomyosin contraction and a self-propelled Voronoi model that was calibrated against the experimental data and included both traction persistence and alignment. Together, the experiments and model reveal how the interplay between persistence and alignment affects the size of a collectively moving cell pack.

\section{Methods}

\subsection{Cell Culture and Sample Preparation}

Madin-Darby Canine Kidney (MDCK) type II cells were used in the experiments. In all experiments except those with SiR-actin (described in the Results section), the cells used were stably transfected with green fluorescent protein (GFP) in the nucleus as described previously \cite{notbohm2016cellular}. Cell monolayers were micropatterned into 1.5 mm islands on polyacrylamide substrates of Young's modulus 6 kPa and thickness of 150 $\mu$m as described previously \cite{saraswathibhatla2020tractions} and detailed in the Supplemental Methods. 

\subsection{Microscopy}

Time lapse microscopy was performed using an Eclipse Ti microscope (Nikon, Melville, NY) with a 20$\times$ numerical aperture 0.5 or 40$\times$ numerical aperture 1.15 objective (Nikon) and an Orca Flash 4.0 camera (Hamamatsu, Bridgewater, NJ) running Elements Ar software (Nikon). For traction experiments, after the imaging, cells were removed from the polyacrylamide substrates by incubating in 0.05\% trypsin for 20 min, and images of the fluorescent particles were collected; these images provided a traction-free reference state for computing cell-substrate tractions. Fixed imaging of stress fibers used an A1R+ confocal microscope with a 40$\times$ NA 1.15 water-immersion objective with a step size of 0.5 $\mu$m using Elements Ar software (all Nikon).

\subsection{Single-Cell Velocity and Traction Measurements}

Cell velocities were measured by using methods of particle tracking \cite{crocker1996methods}. Images of GFP-labeled cell nuclei were segmented using StarDist \cite{schmidt2018cell}. At each time point, every cell nucleus was mapped to the nearest cell nucleus at the next time point using the knnsearch function in Matlab, allowing for cell velocities and trajectories to be computed. 
Cell-induced displacements of the fluorescent particles were measured using Fast Iterative Digital Image Correlation \cite{bar2015fast} using $32\times32$ pixel subsets centered on a grid with a spacing of $8$ pixels ($2.6$ $\mu$m). Tractions were computed using unconstrained Fourier transform traction microscopy \cite{butler2002traction} accounting for the finite substrate thickness \cite{del2007spatio, trepat2009physical}. To quantify the net traction applied by each cell, a Voronoi tessellation was constructed using the center of each cell nucleus, enabling the traction at each grid point to be mapped to each cell. A vector sum of the traction within each cell was computed, giving the net traction applied by each cell. 

\subsection{Numerical Model}
Numerical simulations were performed with the Self-Propelled Voronoi (SPV) model \cite{bi2016motility,barton2017active} implemented in the SAMoS package \cite{sknepnek2017samos}. In the SPV, a cell monolayer is modeled as the Voronoi tiling of the plane with cell centroids acting as Voronoi seeds. We recall that a Voronoi tessellation is a polygonal tiling of the plane based on distances to a set of seed points. For each seed there is a corresponding polygon that contains the region of the plane closer to that seed than to any other. The assumption of Voroni tiling is essential to ensure that smooth motion of cell centroids corresponds to smooth deformations of the cells \cite{barton2017active}.

The state of cell $i$ is described by the position vector, $\mathbf{r}_i$ of its centroid and the direction of self-propulsion (i.e., traction), described by a unit-length vector $\mathbf{n}_i$. The traction vector can be written in terms of the angle $\theta_i$ between $\mathbf{n}_i$ and the $x-$axis of the laboratory reference frame as $\mathbf{n}_i=\left(\cos\theta_i,\sin\theta_i\right)$. The motion of each cell is described by two equations, one for the cell's position $\mathbf{r}_i$ and one for $\theta_i$, 
\begin{eqnarray}
  \mu\dot{\mathbf{r}}_i & = & -\nabla_{\mathbf{r}_i}E + f_0\mathbf{n}_i , \label{eq:pos_pot} \\
  \dot{\theta}_i & = & \frac{1}{\tau_a}\sum_{j}\sin\left( \theta_j-\theta_i \right) + \eta_i, \label{eq:pos_ang}
\end{eqnarray}
where $\mu$ is the friction coefficient that measures dissipation with the environment, $f_0$ is the magnitude of the self-propulsion force in the direction of the traction vector $\mathbf{n}_i$. Variable $\tau_a$ represents the characteristic time required for cell $i$ to align its traction direction with that of its neighbors; smaller $\tau_a$ indicates greater strength of alignment of neighboring tractions.
The variable $\eta_i$ describes a Gaussian random process with $\langle\eta_i\rangle = 0$ and $\langle\eta_i\left(t\right)\eta_j\left(t^\prime\right)\rangle = \tau_p^{-1}\delta_{ij}\delta\left(t-t^\prime\right)$, 
with $\tau_p$ being the characteristic persistence time of the direction of cell traction, and $\langle\dots\rangle$ denoting an average with respect to the distribution of the realizations $\eta_i$.
The $j$-sum is over all neighbors of cell $i$ within a cutoff distance $r_{cut}$, set to be equal to the average cell diameter. 
Finally, 
\begin{equation}
    E = K_A\sum_i \left(A_i - A_0\right)^2 + \Gamma_P\sum_i\left(P_i - P_0\right)^2, \label{eq:VM}
\end{equation}
is the energy functional of the vertex model \cite{farhadifar2007influence}. $K_A$ is the area modulus, i.e., the energy cost of the cell $i$ having area $A_i$ that is different from the preferred area $A_0$. $\Gamma_P$ is the perimeter modulus, i.e., the energy cost of cell $i$ having perimeter $P_i$ that is not equal to the preferred perimeter $P_0$. We note that in Eq.\ (\ref{eq:pos_pot}), the gradient is computed using the position of the cell centroid $\mathbf{r}_i$, while the energy functional in Eq.\ (\ref{eq:VM}) is typically written in terms of positions of the vertices that are meeting points of three or more cell junctions. If one restricts to the Voronoi tiling, as is the case in the SPV, there is a one-to-one mapping between positions of cell centroids and vertices, which makes it possible to directly compute the force on the cell centroid from Eq.\ (\ref{eq:pos_pot}). The calculation is tedious but straightforward \cite{barton2017active}. Cell velocities in the model are computed by taking the displacement at different time points and dividing by time, which is consistent with the method used to compute velocities in the experiments.

\section{Results}

\begin{figure}
    \centering
    \includegraphics[width=0.5\linewidth]{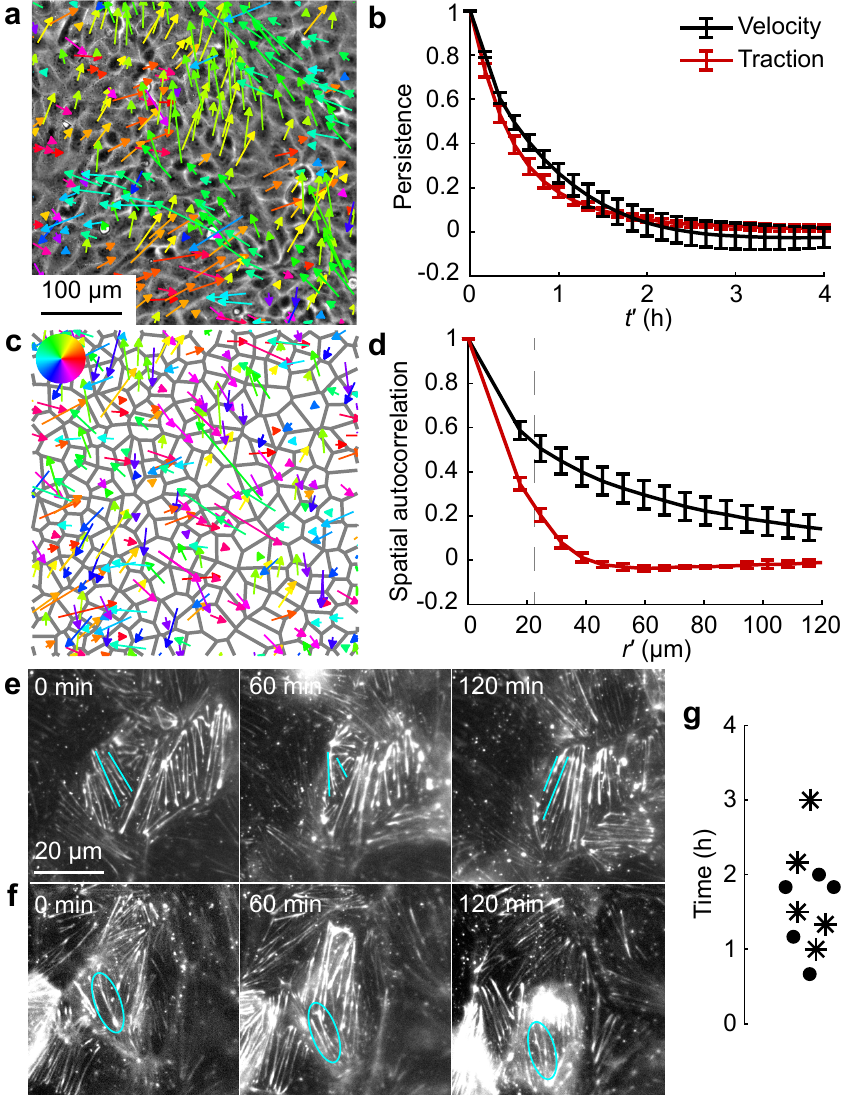}
    \caption{Temporal persistence and spatial correlation of velocity and traction. (a) Cell velocities (arrows) overlaid on an image of a cell monolayer. Colors indicate the angle of the cell velocity  (see color circle). 
    (b) Persistence, defined as the temporal autocorrelation, of single-cell velocity and traction. In all figures, lines with error bars represent mean values and standard deviations over at least four cell monolayers and 6 h of imaging. (c) Single-cell tractions overlaid on a Voronoi tessellation of a cell monolayer. Colors indicate the angle of traction (see color circle).
    (d) Spatial autocorrelation of single-cell velocity and traction. In all figures, vertical dashed lines represent the average cell diameter.
    (e) Representative images of stress fibers reorienting over time. Solid lines indicate the orientation of a pair of stress fibers at different time points. (f) Representative images of a stress fiber disappearing over time (circled). (g) Time required for stress fibers to reorient (circles) or disappear (stars).}
    \label{fig1}
\end{figure}

To begin, we seeded MDCK cells in confluent monolayers, imaged them over time, and tracked the nuclei \cite{schmidt2018cell, crocker1996methods} to measure each cell's velocity (Fig.\ \ref{fig1}a). The persistence of cell velocity was defined by the temporal autocorrelation, $C(t^\prime) = \langle \hat{\mathbf{v}}\left({t}\right) \cdot \hat{\mathbf{v}}\left({t}-{t}^\prime\right) \rangle $, where $\hat{\mathbf{v}}({t})$ is the unit vector corresponding to the direction of cell velocity at time ${t}$, and $\langle ... \rangle$ indicates an average over all cells in all pairs of time points that are $t^\prime$ apart. To quantify the persistence time, we used the time over which the persistence decreased to a value of $0.2$, {which corresponds to the time required for the direction of cell velocity to change by 78$^\circ$}. The persistence time was $\approx1.5$ h (Fig.\ \ref{fig1}b), consistent with prior research \cite{henkes2020dense}.
In addition to persistence, the cell velocity fields showed packs of cells moving in the same direction (Fig.\ \ref{fig1}a; Supplemental Video 1). The pack size was quantified by a spatial autocorrelation of $\hat{\mathbf{v}}$, which showed correlated cell velocities over distances of several cell diameters (Fig.\ \ref{fig1}d). As with the persistence time, the pack size was quantified by the distance over which the correlation dropped to $0.2$, which was $\approx90$ $\mu$m (Fig.\ \ref{fig1}d), setting a typical pack to have the size of $\approx180$ $\mu$m ($\approx8$ cell diameters), which matches the order of magnitude of previous measurements of velocity correlations in confluent epithelial sheets \cite{angelini2011glass, garcia2015physics, loza2016cell, malinverno2017endocytic, henkes2020dense}.

To relate these temporal and spatial velocity correlations to the tractions at the cell-substrate interface, we used traction force microscopy \cite{bar2015fast, dembo1999stresses, butler2002traction, del2007spatio, trepat2009physical}, which gave a grid of traction measurements between the cell monolayer and the substrate spaced uniformly by $2.6$ $\mu$m. As cells are not uniformly spaced, the traction data did not map directly to each cell. To resolve this issue, we produced a Voronoi tessellation approximating the cell outlines (Supplemental Fig.\ S1) \cite{kaliman2016limits}, which enabled an approximate mapping between each cell and its traction (Supplemental Fig.\ S2a). Then, the vector sum of traction applied by each cell was computed, yielding the net traction produced by that cell (Fig. 1c). {To verify the accuracy of the net traction computed by the Voronoi-based estimate of the cell outlines, we compared the Voronoi-based net traction to the net traction computed from segmenting each cell, with results showing close agreement (Supplemental Fig.\ S2b, c). Therefore, we used the Voronoi-based estimate of the cell outlines for the remainder of our study.}
The magnitude of net traction applied by each cell fluctuated in time but rarely reached $0$ (Supplemental Fig.\ S2d). Importantly, the direction of net traction coincided closely with the orientation of stress fibers in cells, consistent with the notion that stress fibers generate tractions (Supplemental Fig.\ S2e-f). The noise floor of this new measurement was quantified by computing the net traction applied by isolated cells, which should be zero as inertial forces are negligible \cite{maruthamuthu2011cell, tanimoto2014simple}. The average was $10$ Pa (Supplemental Fig.\ S2g), which defines the experimental noise floor, and, importantly, was nearly an order of magnitude smaller than the average net traction applied by cells in a monolayer, $\approx80$ Pa. 

The net traction is not merely a high resolution measurement of tractions, but rather a measurement of the total force applied by each cell to the substrate. As such, it enables new investigation, for example, of the temporal persistence and spatial correlation of each cell's traction (Fig.\ \ref{fig1}b, d).
The traction persistence time, defined as the time over which the correlation decreased to $0.2$, was found to be $\approx1.5$--$2$ h (Fig.\ \ref{fig1}b, Supplemental S2h). To investigate the cytoskeletal origins of the traction persistence time, we performed live fluorescence imaging of F-actin labeled with SiR-actin \cite{lukinavivcius2014fluorogenic}. After observing the temporal evolution of stress fibers in multiple cells, we classified our observations into two cases. In the first, the stress fibers reoriented with time (Fig.\ \ref{fig1}e); in the second, the stress fibers disappeared after some time (Fig.\ \ref{fig1}f). We then quantified the time required for stress fibers to reorient or disappear, with times averaging $1.5 \pm 0.5$ and $2.0 \pm 0.7$ h, respectively (Fig.\ \ref{fig1}g). As these times match the persistence times of traction, these data provide evidence that the measurement of net traction is accurate and that its temporal dynamics are physically meaningful. 
Turning now to the spatial correlation of traction, if tractions applied by neighboring cells were fully uncorrelated, the autocorrelation would decay to zero over a distance of approximately one cell diameter ($\approx22$ $\mu$m). By contrast, the averaged autocorrelation decayed to zero over a distance of two cell diameters, indicating that cells coordinate with their neighbors to apply traction in the same direction (Fig.\ \ref{fig1}d). Hence, cell tractions are correlated in both time and space.

\begin{figure}
    \centering
    \includegraphics[width=0.5\linewidth]{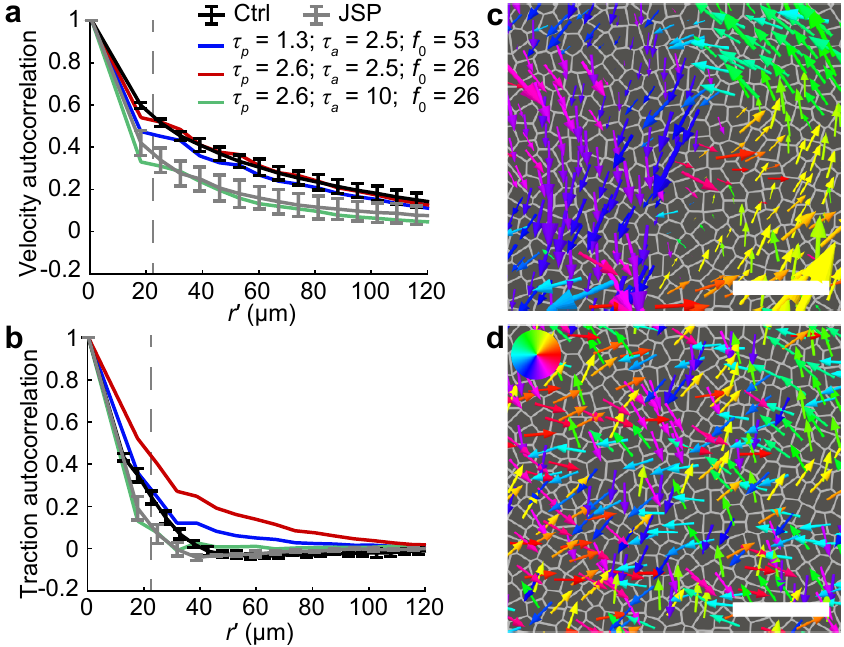}
    \caption{Evidence for traction alignment spanning multiple cells. 
    (a, b) Spatial autocorrelation of cell tractions (a) and velocities (b) in experiments and model. Values in the legend in panel a correspond to different traction persistence and alignment times, $\tau_p$ and $\tau_a$, measured in hours, and force $f_0$ measured in nN.
    (c, d) Representative images of cell velocities (c) and net cell tractions (d) in the model, for $\tau_p = 1.3$ h, $\tau_a = 2.5$ h and $f_0 = 53$ nN.
    {Colors indicate the angle of velocity or traction (see color circle).}
    The scale bars correspond to $100 \:\mu$m.
    }
    \label{fig2}
\end{figure}

We next sought to identify the relative contributions of traction persistence and correlation on the size of a collectively moving cell pack. To begin, we considered temporal persistence, as prior models have suggested that increasing persistence increases the pack size \cite{bi2016motility, henkes2020dense, mitchel2020primary}. In an effort to alter actin dynamics, we used the F-actin stabilizer jasplakinolide (JSP) \cite{holzinger2009jasplakinolide}, which increased traction persistence and decreased traction magnitude by factors of $\approx2$ (Supplemental Fig. S3a-c). Even though the JSP increased traction persistence, it reduced the spatial correlation of velocity (Fig.\ \ref{fig2}a), in contrast to our expectation. A potential resolution is that the treatment also {caused a statistically significant reduction in} the spatial correlation of traction  (Fig. 2b, {Supplemental Fig.\ S3d, e}), suggesting that the velocity correlation may be determined more so by the traction correlation. 
Other chemical treatments produced similar results: using ML 141 to inhibit the front-back polarization molecule Cdc42 \cite{cau2005cdc42} and altering actin polymerization with cytochalasin D \cite{casella1981cytochalasin} also increased the persistence time {but decreased the spatial correlation of traction. As with JSP, both ML 141 and cytochalasin D decreased the velocity correlation, though these treatments also reduced the magnitude of traction produced (Supplemental Fig.\ S3). Together, these data suggest that traction correlation may have a larger effect than traction persistence on the size of a collectively moving cell pack.}
An alternative explanation is that the treatments affected pack size by altering the distance over which cell--cell forces propagate. To consider this explanation, we quantified propagation of tensile stresses in the monolayer by using monolayer stress microscopy \cite{tambe2011collective, tambe2013monolayer, saraswathibhatla2020spatiotemporal} and computing an autocorrelation. Results showed that the correlation length of tension was not statistically affected by treatment with JSP (Supplemental Fig. S4), which ruled out the explanation based on cell--cell forces, {thus supporting the idea that the spatial correlation of traction had the stronger effect on pack size.}

{A potential confounding factor in the experimental data is that the traction magnitude changed in addition to the traction persistence and correlation (Supplemental Fig.\ S3). Experiments alone cannot resolve this issue, because the three variables (traction magnitude, persistence, and correlation) cannot be controlled independently in the experiments. Therefore, we turned to the SPV model} \cite{bi2016motility, barton2017active, sknepnek2017samos}. 
Although the original implementation of the SPV model \cite{bi2016motility} included only temporal persistence of traction, our experimental data give evidence for both persistence and alignment. Therefore, we {extended the SPV model by adding} a term in Eq. \ref{eq:pos_ang} for alignment of traction between neighboring cells, similar to recent studies \cite{barton2017active, lin2021energetics}.
The SPV model includes sufficient cell-level detail to {estimate, to order of magnitude, the} parameters of the experimental system \cite{henkes2020dense}. {Briefly, the estimates are as follows. The unit of length was chosen to match the average cell diameter in the experiments, $\ell=22$ $\mu$m, which set the the preferred cell area $A_0=\pi\ell^2/4$ in the SPV model. Furthermore,} $A_0$ can be combined with {the preferred cell perimeter} $P_0$ into a single dimensionless parameter, $p_0=P_0 A_0^{-1/2}$, which is one factor controlling whether the collective cell behavior in the model is liquid- or solid-like \cite{bi2015density, bi2016motility}. {In the SPV model simulations}, we used an input value of $p_0=3.8$, which is in the solid regime but near the border to a liquid. Adding self-propulsion then raises the actual values of $PA^{-1/2}$ above the input value of $3.8$, consistent with measured values of $PA^{-1/2}$ in MDCK cells \cite{saraswathibhatla2020tractions} and with the fluid-like migration observed in our experiments (Supplemental Video 1).
We used the stiffness parameters $K$ and $\Gamma$ estimated in ref.\ \cite{henkes2020dense}. 
The value of $f_0$ was estimated from the mean magnitude of net traction, $110$ Pa (Supplemental Fig.\ S3a), multiplied by the average cell area, $22\times22$ $\mu$m$^2$ to give 53 nN. The friction factor $\mu$ was tuned to achieve a match between cell migration speeds in the experiments and model (average speed of $18$ $\mu$m/h, Supplemental Fig.\ S5); a value of $\mu = 265$ pN$\cdot$h/$\mu$m 
gave the best match and is of the same order of magnitude as prior estimates \cite{duclos2017topological, henkes2020dense}.
Earlier references wrote the equations using variable $v_0=f_0/\mu$ in place of $f_0$, with $v_0$ having units of velocity \cite{bi2016motility, barton2017active}. Here, we chose to use $f_0$ due to its connection to the single-cell net traction measured in the experiments.

{There remained two free parameters in the model,} the persistence and alignment time scales, $\tau_p$ and $\tau_a$, respectively. We {used a value of} $\tau_p = 1.3$ h to match the experimentally measured traction persistence time (Fig.\ \ref{fig1}b). {We then adjusted $\tau_a$ to achieve a good fit to the traction correlations in the untreated control data, finding a good match for $\tau_a=2.5$ h (Fig.\ \ref{fig2}b). Importantly, using the value $\tau_a=2.5$ h, the model also matched closely to the velocity correlation data (Fig.\ \ref{fig2}a; see the Supplemental Note for a comparison of experimental and simulated velocity persistences).} The matched value of $\tau_a = 2.5$ h is slightly longer than the $1.5$--$2$ h required for stress fibers to reorient in a cell (Fig.\ \ref{fig1}g), which is reasonable given that the multicellular realignment of traction associated with $\tau_a$ could not be faster than the time required for each cell to reorganize its own stress fibers. 
{Representative images of cell velocity and traction in the model (Fig.\ \ref{fig2}c,d) show collective cell packs and neighboring cells applying traction in the same direction, which is reminiscent of the experimental data.}

{With the model parameters chosen to match experimental control conditions, we then adjusted the parameters to mimic the experiments with JSP. To this end,} we increased the traction persistence time $\tau_p$ and decreased the traction magnitude $f_0$ by factors of two, {consistent with the experimental data (Fig.\ S3a--c)}. The effect of this change on the velocity correlation was small (Fig.\ \ref{fig2}a), suggesting that $\tau_p$ alone had a relatively small effect on the velocity correlations. {Moreover, these altered parameters increased the spatial correlations in velocity and traction}, opposite to the experimentally observed effects of JSP (Fig.\ \ref{fig2}a, b). Matching the experimental data required {us to increase} both $\tau_p$ and $\tau_a$, which indicates that the multicellular traction alignment is essential to capture the experimental data. {We performed a more thorough theoretical study of the combined effects of $\tau_p$ and $\tau_a$} by decreasing and increasing $\tau_p$ by factors of two and subsequently varying $\tau_a$ through a range of values (Supplemental Fig.\ S5). Consistent with prior models \cite{bi2016motility, henkes2020dense, mitchel2020primary}, increasing $\tau_p$ caused larger collective packs in the absence of multicellular traction alignment (i.e., $\tau_a\to\infty$). Additionally, for larger values of $\tau_p$, less traction alignment (i.e., larger $\tau_a$) was required to match the experimental data (Supplemental Fig.\ S5), indicating that increasing either single-cell persistence or multicellular alignment increases the size of a collectively moving cell pack. Importantly, for all values of $\tau_p$ studied, a finite $\tau_a$ was required to match {the simulations to the} experimental data. {In the absence of traction alignment, the pack size was notably smaller than the experimental values, typically by a factor of $\approx 4$ (Supplemental Fig.\ S5). Hence, the model demonstrates that alignment of traction between neighboring cells has a strong effect on the pack size.}

\begin{figure}
    \centering
    \includegraphics[width=0.5\linewidth]{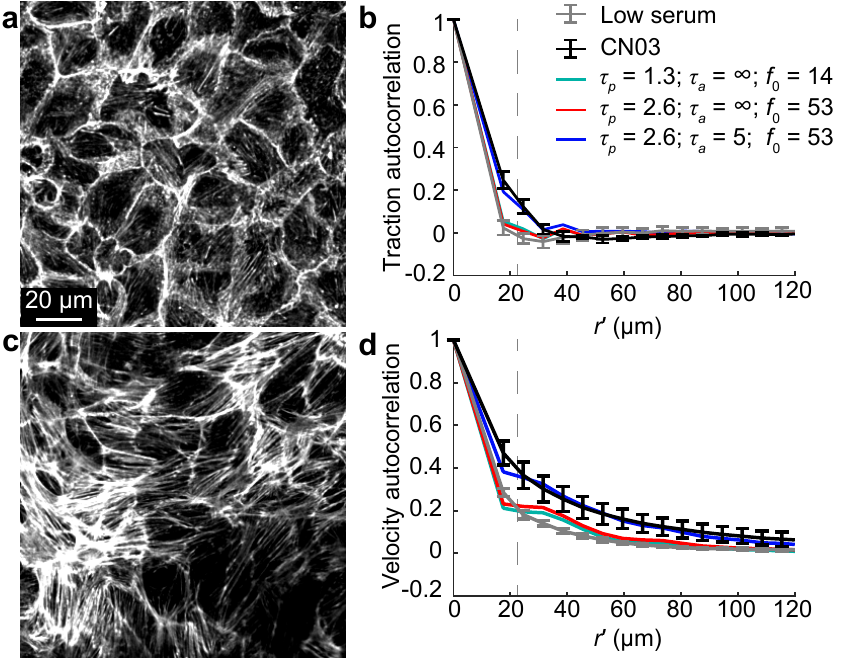}
    \caption{Effect of stress fiber activity on cell-cell traction alignment and collective pack size. (a) Image of F-actin in low serum conditions. (b) Traction autocorrelation in experiments and model. (c) Image of F-actin in CN03-treated cells. (d) Velocity autocorrelation in experiments and model. In panels b and d, $\tau_p$ and $\tau_a$ are measured in hours; $f_0$ is measured in nN. }
    \label{fig3}
\end{figure}

The experiments and model indicate that cells coordinate their tractions with their neighbors, which in turn affects the size of a collectively moving pack. To explore this finding further, we wondered if it would be possible to turn off the correlation of traction between neighbors, thereby reducing the pack size. To this end, we decreased contractility by culturing the cells in low (1\%) serum for 24 h. Under these conditions, cells expressed few stress fibers (Fig.\ \ref{fig3}a), and the traction autocorrelation dropped nearly to zero at one cell diameter (Fig.\ \ref{fig3}b), suggesting essentially no coordination of traction between neighbors. {Importantly, even in low serum conditions, the temporal persistence of traction stayed well above the imaging period of 10 min (Supplemental Fig. S6b), indicating that the traction data remained above the noise floor of the experiments. Hence, the loss of spatial correlation was not an artifact due to noise but rather an indicator of reduced coordination between neighboring cells.}
Addition of the Rho activator CN03 (3 $\mu$g/mL) reversed the effects of low serum, causing more pronounced stress fibers that aligned between neighboring cells (Fig.\ \ref{fig3}c) and increasing the traction autocorrelation such that it decayed to zero over a distance of $\approx2$ cell diameters (Fig.\ \ref{fig3}b). As expected, the velocity autocorrelation was smaller for serum-starved compared to CN03-treated conditions (Fig.\ \ref{fig3}d). To confirm that the low serum conditions essentially turned off the multicellular traction alignment, we returned to the model. {To mimic the low serum conditions, we switched off traction alignment by setting $\tau_a\to\infty$, which resulted in a good match with the experimental traction and velocity correlations.} To match the CN03-treated conditions, we noted that the CN03 increased net traction magnitude and traction persistence by factors of approximately $4$ and $2$, respectively (Supplemental Fig.\ S6). Therefore, {in the simulations,} we {respectively} increased $f_0$  and $\tau_p$ by factors of $4$ and $2$ but retained no multicellular traction alignment (i.e., $\tau_a\to\infty$). Despite increasing the cell persistence time $\tau_p$,  {the simulations did not reveal any} significant change in velocity autocorrelation. Instead, a match {between the model and} the CN03-treated {experimental} data required that we use a finite traction alignment time, $\tau_a = 5$ h (Fig.\ \ref{fig3}b, d). Hence, the experiments {and simulations confirm} the multicellular traction alignment and demonstrate that it results from alignment of stress fibers between neighboring cells.

\section{Conclusions}

Here, we have demonstrated that cells in a collective align their cell-substrate tractions with those of their neighbors, {which leads to complex large-scale patterns in collective motion that are reminiscent of the rich collective behaviors observed in flocks of birds or schools of fish. For birds and fish, the collective alignment of propulsion results from visual cues. For epithelial cells, the underlying causes of the alignment are not yet clear, but} it is likely that the alignment mechanism involves a combination of biochemical signalling, e.g., via correlation of front-back polarity, which can be coordinated between neighboring cells \cite{das2015molecular, hino2020erk}, and mechanical interaction.
{Numerical simulations of the SPV model revealed that the collective motion resulting from the complex physics and biological signaling can be captured quantitatively by a relatively simple model. Our results, therefore, show that even without knowledge of the specific mechanisms for traction alignment, the complex motion patterns in cell monolayers can be explained within the framework of physics of active matter.} This is in line with the growing notion that mechanical cues play an equally important role as chemical signaling in controlling collective cell behavior at tissue scales. More broadly, by increasing the size of a collective cell pack, the multicellular traction alignment tempers randomness in cell motion caused by biological variability, which could be important for formation and repair of tissues.

\section*{Acknowledgments}
We thank Christian Franck for use of the 40$\times$ microscope objective. A.S., E.E.G., and J.N. were supported by National Science Foundation grant number CMMI-1660703 and the University of Wisconsin-Madison Office of the Vice Chancellor for Research and Graduate Education with funding from the Wisconsin Alumni Research Foundation. R.S. acknowledges support by the UK BBSRC (Award BB/N009789/1). S.H. acknowledges support by the UK BBSRC (grant number BB/N009150/1-2).

\bibliographystyle{elsarticle-num} 
\bibliography{cell_migration}

\end{document}


\renewcommand{\theequation}{S\arabic{equation}}

\begin{center}

\Large \textbf{Supplementary Information: Coordinated tractions increase the size of a collectively moving pack in a cell monolayer}

\normalsize
Aashrith Saraswathibhatla,$^1$
Silke Henkes,$^2$
Emmett E. Galles,$^1$
Rastko Sknepnek,$^{3,4}$
Jacob Notbohm$^1$

$^1$Department of Engineering Physics, University of Wisconsin-Madison, Madison, Wisconsin 53706, USA\\
$^2$School of Mathematics, University of Bristol, Bristol BS8 1TW, United Kingdom\\
$^3$School of Science and Engineering, University of Dundee, Dundee DD1 4HN, United Kingdom\\
$^4$School of Life Sciences, University of Dundee, Dundee DD1 5EH, United Kingdom

\end{center}

Correspondence should be sent to: jknotbohm@wisc.edu

\vspace{0.2in}

\section{Supplemental Figures}

\vspace{11pt}
\begin{center}
\includegraphics[width=3.5in]{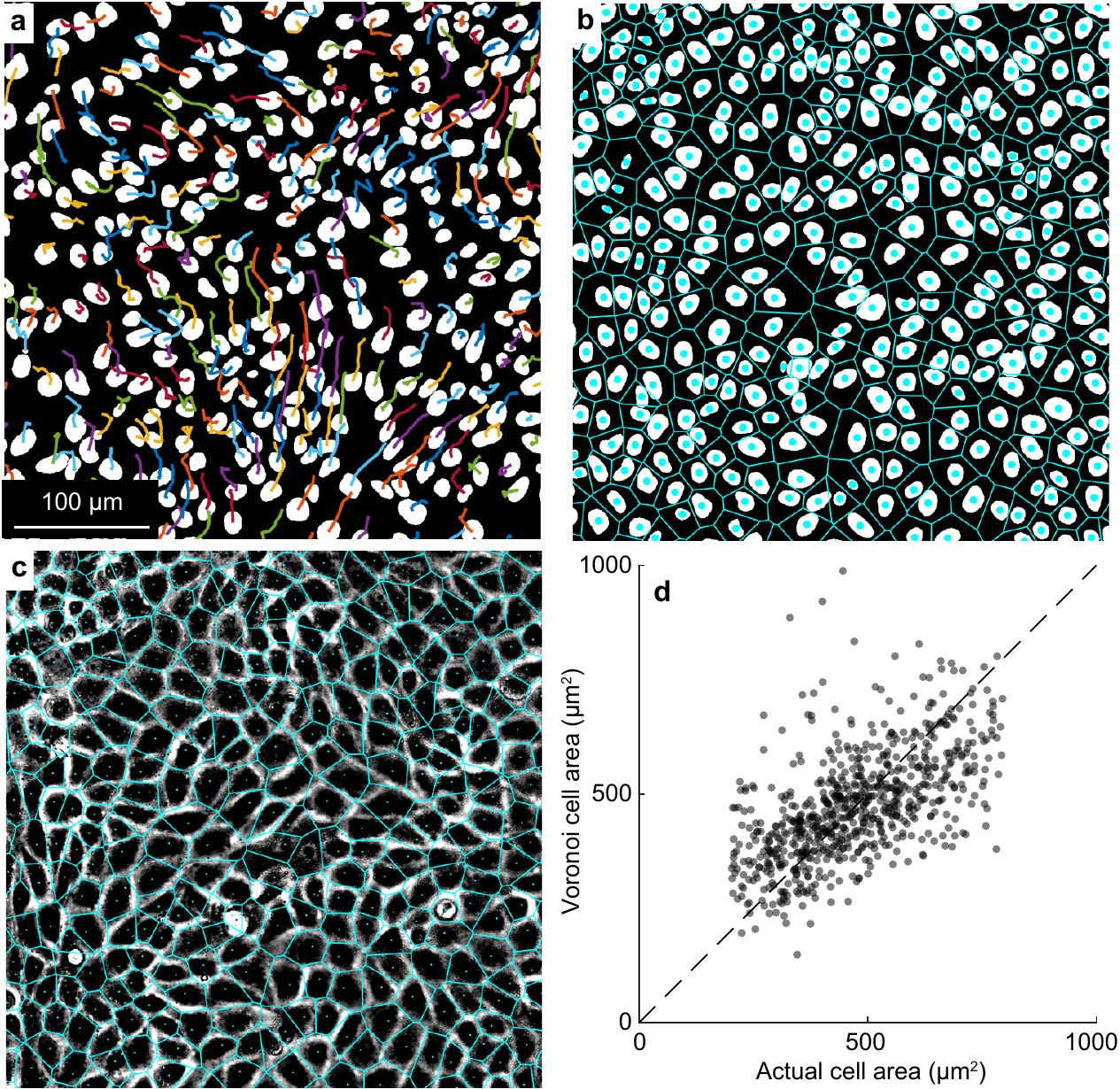}
\end{center}
\vspace{-12pt}
\textbf{Figure S1:} Tracking each cell and its outline in a cell monolayer. (a) Cellular trajectories built by tracking each cell’s nucleus which is labeled with green fluorescent protein. \textcolor{black}{Colors are chosen randomly to help distinguish between neighboring trajectories.} (b) Voronoi tessellation built using the nuclei centroids. (c) Overlay of the Voronoi tessellation on a phase contrast image of the corresponding cell monolayer. (d) Scatter plot comparing cell areas from the Voronoi tessellation and the actual cell areas. Each dot corresponds to a different cell.

\begin{center}
\includegraphics[width=5.75in]{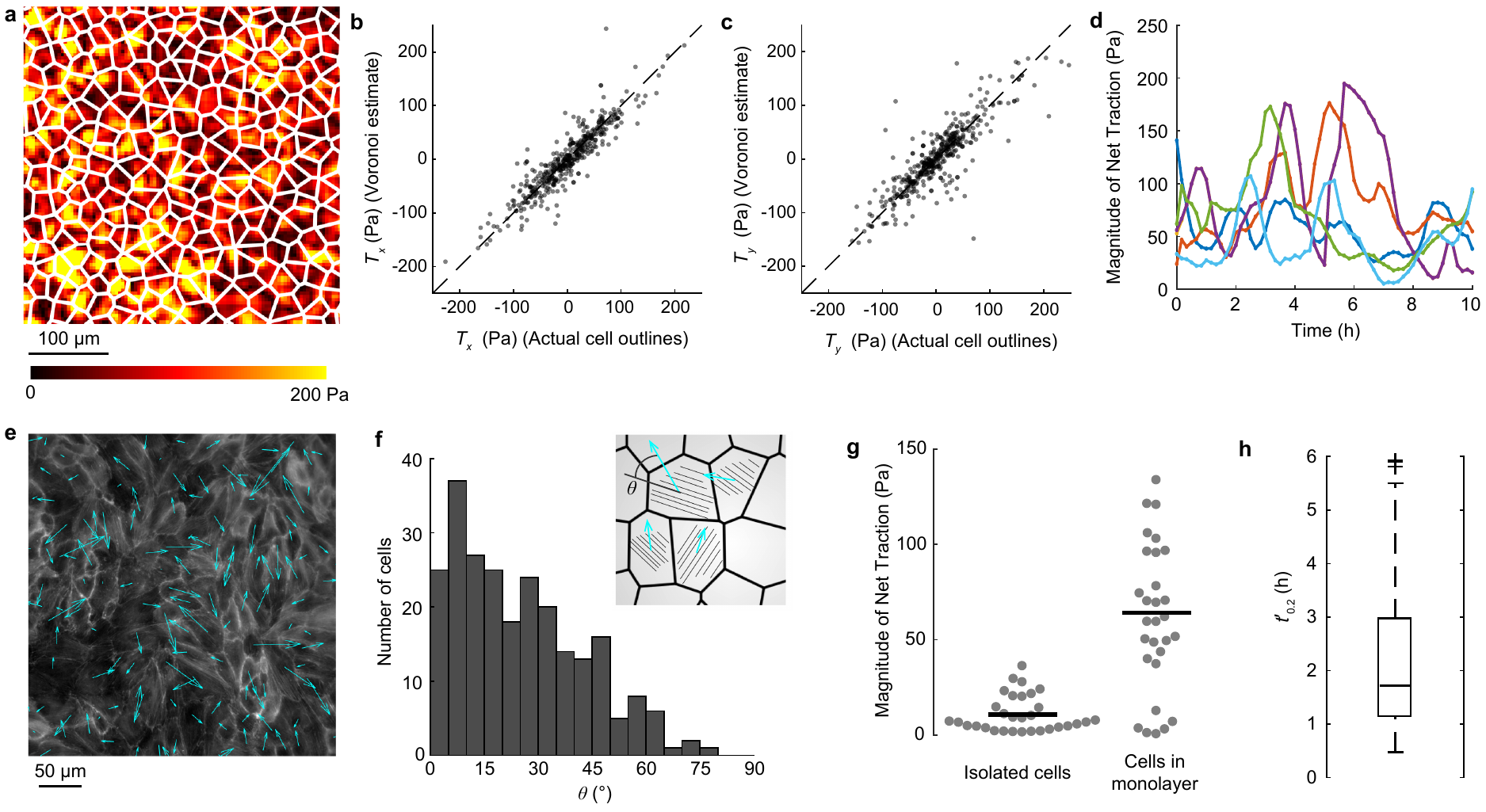}
\end{center}
\vspace{-14pt}
\textbf{Figure S2:} Net traction of cells in a monolayer.
(a) Overlay of Voronoi tessellation on traction color map.
(b, c) Comparison between $x$ (b) and $y$ (c) components of net traction computed by Voronoi approximation of the cell outline and the actual cell outline.  Each dot corresponds to a different cell.
(d) Magnitude of vector sum of traction (i.e., net traction) applied by five representative cells (colors) over time. 
(e) Images of F-actin labeled with phalloidin. Tractions were measured simultaneously (see materials and methods below); the arrows show the net traction applied by each cell.
(f) Histogram of angle $\theta$ between the orientation of stress fibers and direction of traction for different cells. The inset defines angle $\theta$ where stress fibers are shown by the black lines and traction directions are shown by the cyan arrows.
(g) Magnitude of net traction for isolated cells and cells in a monolayer. Each dot represents a different cell. As isolated cells produce zero net traction, the data for isolated cells quantifies the noise floor of the measurement. The magnitude of net traction applied by cells in a monlayer is statistically larger than that applied by isolated cells ($p < 0.001$, rank sum test). 
(h) Box plot of $t^\prime_{0.2}$, the time at which the temporal persistence of traction reaches a value of 0.2, for $\approx600$ cells.

\begin{center}
\includegraphics[width=6.5in]{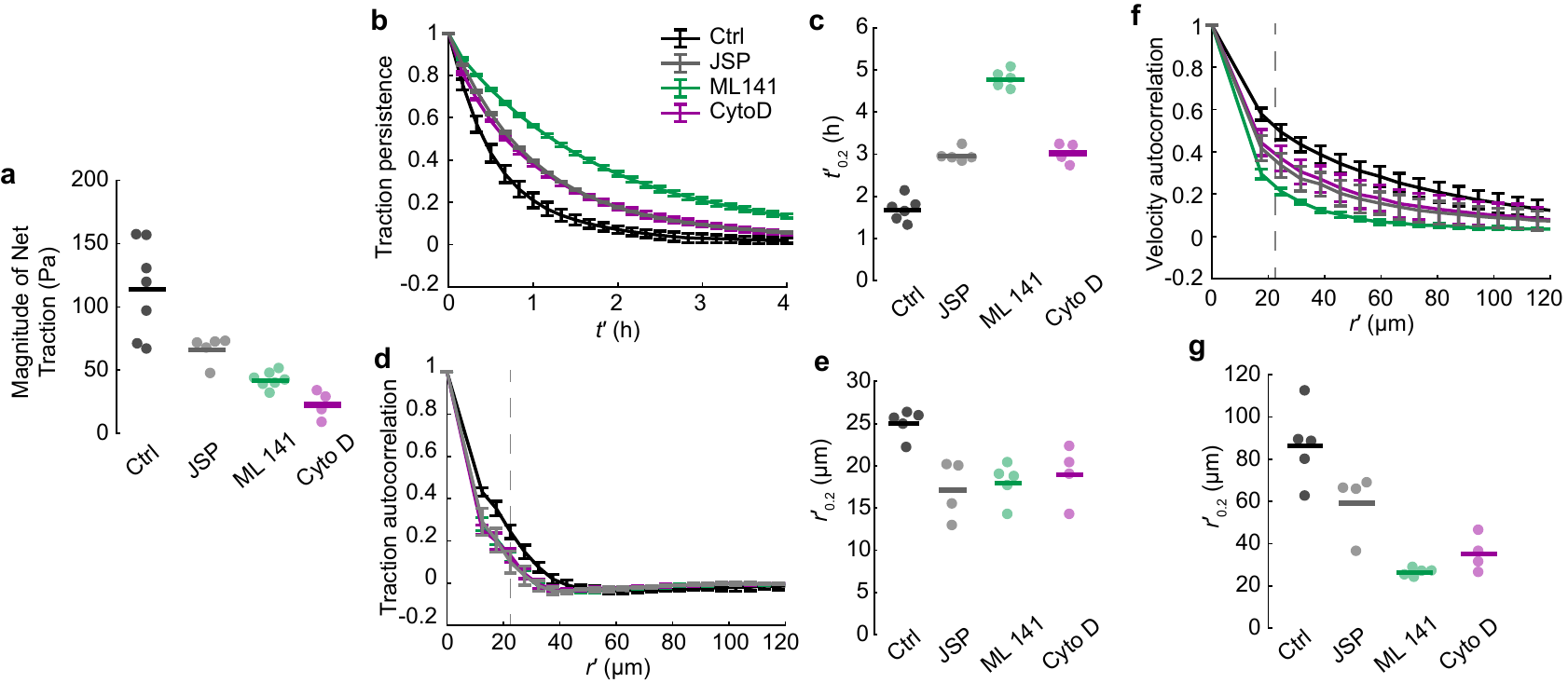}
\end{center}
\vspace{-16pt}
\textbf{Figure S3:} Perturbing traction temporal persistence and spatial correlation.
(a) Magnitude of net traction after treating cell monolayers with Jasplakinolide (JSP), ML 141, or cytochalasin D (Cyto D). All treatments were statistically different from control (Ctrl) ($p < 0.01$, ANOVA with Tukey correction for multiple comparisons).  
(b) Average temporal persistence of traction in response to the treatments.
(c) Temporal persistence time of traction. All treatments were statistically different from control ($p < 0.01$, ANOVA with Tukey correction). 
(d) Average spatial correlation of traction in response to the treatments.
(e) Correlation length of traction. All treatments were statistically different from control ($p < 0.01$, ANOVA with Tukey correction). 
(f) Spatial autocorrelation of velocity in response to the treatments.
(g) Correlation length of velocity. All treatments were statistically different from control ($p < 0.01$, ANOVA with Tukey correction). 
In panels a, c, e, and g each dot represents an average from a different cell monolayer. 

\vspace{11pt}
\begin{center}
\includegraphics[width=3.75in]{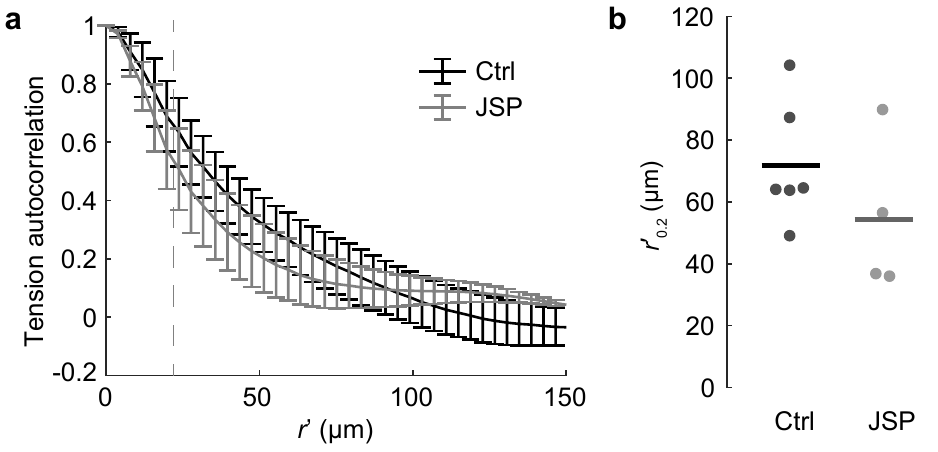}
\end{center}
\vspace{-12pt}
\textbf{Figure S4:} Effect of Jasplakinolide (JSP) on tension correlation length. (a) Monolayer tension, defined as the mean principal stress, was computed using monolayer stress microscopy and an autocorrelation was performed. (b) Correlation length of tension $r^\prime_{0.2}$ from control cell monolayers and monolayers treated with JSP. The data are not significantly different by rank sum test. Each dot represents a different cell monolayer; horizontal lines show means.

\vspace{11pt}
\begin{center}
\includegraphics[width=7.1in]{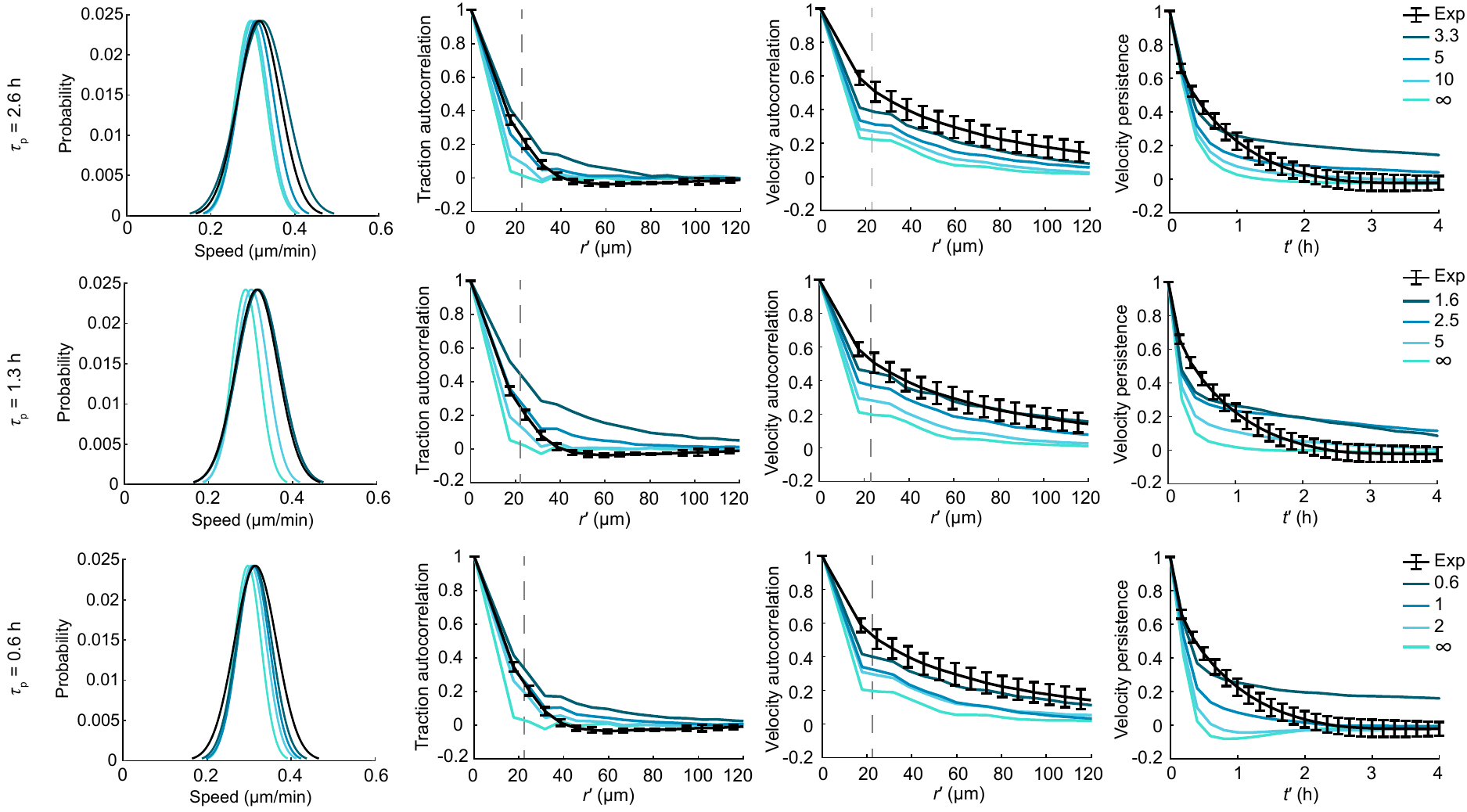}
\end{center}
\vspace{-12pt}
\textbf{Figure S5:} Combined effects of persistence and alignment times, $\tau_p$ and $\tau_a$. Histograms of cell speed and graphs of traction autocorrelation, velocity autocorrelation, and velocity persistence are shown for $\tau_p=2.6$, $1.3$, and $0.6$ h. Values in the legends on the right correspond to different traction alignment times $\tau_a$ in units of h. For all plots, $f_0 = 53$ nN.

\newpage
\begin{center}
\includegraphics[width=5.5in]{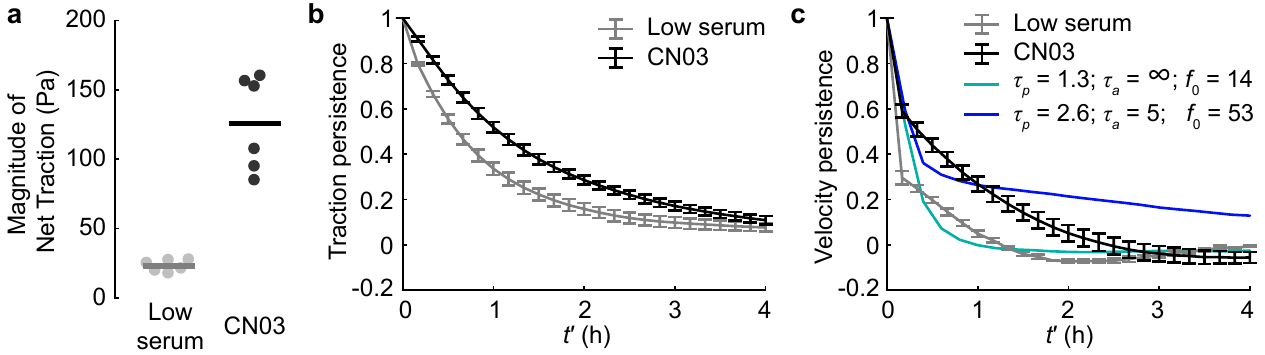}
\end{center}
\vspace{-12pt}
\textbf{Figure S6:} Effects of low serum and Rho-activator CN03. (a, b) CN03 treatment increased traction magnitude (a, $p < 0.001$, rank sum test), and traction persistence (b). (c) Velocity persistence in the experiments and the model. $\tau_p$ and $\tau_a$ have units of h; $f_0$ has units of nN.

\section{Supplemental Video Caption}

\begin{center}
\includegraphics[width=2.2in]{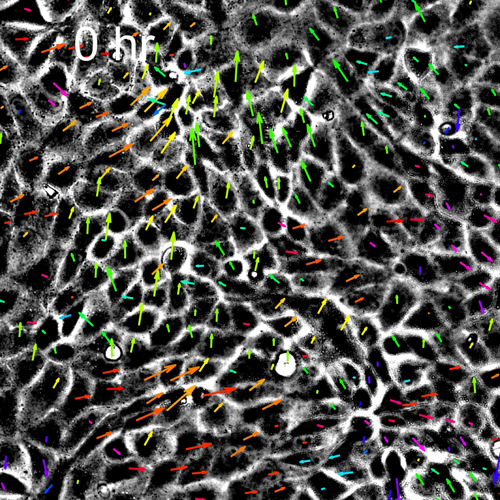}
\end{center}
\vspace{-12pt}
\textbf{Video 1:} Time lapse video of cell velocities (arrows) overlaid on phase contrast images of the cell monolayer. Colors indicate the angle of the cell velocity with respect to the horizontal axis (see color circle in Fig. 1). The size of the field of view is $360 \times 360$ $\mu$m.


\section{Supplemental Note---Persistence of Velocity}

Here, we compare the temporal persistence of velocity between experiments and model. A temporal autocorrelation of cell velocities showed that increasing traction alignment (i.e., decreasing $\tau_a$) caused a greater temporal correlation, indicating more persistent collective motion (Fig.~S5). Compared to the aucorrelations in traction and velocity, the persistence of velocity did not agree as well with the experimental data---at short time scales it decayed faster in the model than in the experiments for studied values of $\tau_a$. Additionally, for long times and small values of $\tau_a$, the velocity persistence in the model was larger than the experiments.  These differences suggest that there are additional factors in the experiments that are not considered in the model. One such factor is likely to be that in MDCK cells waves of collective cell motion oscillate over space with time periods of a few hours \cite{deforet2014emergence, notbohm2016cellular, petrolli2019confinement}, which would reduce the persistence on these time scales. These oscillations were not included in the model, which would explain the larger persistence at long times in the model than in the experiments. 

\section{Materials and Methods}

In this section, we give a full description of all materials and methods used.

\subsection{Cell culture}
Madin-Darby Canine Kidney type II cells were used in the experiments. In all experiments except those with SiR-actin, the cells used were stably transfected with green fluorescent protein (GFP) in the nucleus as described previously \cite{notbohm2016cellular}. The cells were maintained in low-glucose Dulbecco's modified Eagle's medium (12320-032; Life Technologies, Carlsbad, CA) with 10\% fetal bovine serum (Corning, NY) and 1\% G418 (Corning) in an incubator at 37$^\circ$C and 5\% CO2. The cells were passaged every 2 or 3 days, and used within 20 passages, during which no changes to cell morphology, proliferation, or GFP expression were observed. For experiments that treated cells with cytochalasin D, jasplakinolide, and ML-141, and the corresponding control, cell medium was replaced with medium containing 2\% fetal bovine serum 8-12 h  before the start of the experiment. For experiments referred to as low serum, medium was replaced with medium containing 1\% serum 24 h before the start of the experiment.

\subsection{Polyacrylamide substrates and micropatterning}
Cell monolayers were micropatterned on polyacrylamide substrates as described previously \cite{saraswathibhatla2020tractions}. \textcolor{black}{Briefly, we fabricated polyacrylamide gels with Young’s modulus of 6 kPa and thickness of 150 $\mu$m with 0.014\% (weight/volume) fluorescent particles (diameter 0.2 $\mu$m, carboxylate modified; Life Technologies) embedded in the gel. The gels were centrifuged upside down during the gel polymerization to localize the fluorescent particles at the top of the gel.} Next, polydimethysiloxane masks with holes (1.5 mm diameter) were placed on the gels before functionalizing with sulfo-SANPAH and collagen I, thereby constraining the collagen to circular patterns on the gels. 500 $\mu$L of cell solution of concentration 0.5 million cells/mL was pipetted onto the masks and incubated at 37$^\circ$C for 2 h. The masks were removed, and confluent cell monolayers were formed within 10-12 h. 

\subsection{Microscopy}
Time lapse microscopy was performed using an Eclipse Ti microscope (Nikon, Melville, NY) with a 20$\times$ numerical aperture 0.5 objective (Nikon) and an Orca Flash 4.0 camera (Hamamatsu, Bridgewater, NJ) running Elements Ar software (Nikon). Live imaging of stress fibers and the experiments that measured tractions and imaged stress fibers simultaneously used the same microscope and a 40$\times$ numerical aperture 1.15 objective (Nikon). Fixed imaging of stress fibers used an A1R+ confocal microscope with a 40$\times$ NA 1.15 water-immersion objective with a step size of 0.5 $\mu$m using Elements Ar software (all Nikon).

\textcolor{black}{For time lapse imaging, fluorescent and phase contrast images of a region of size 600$\times$600 $\mu$m$^2$ in the center of the 1.5 mm diameter cell island were captured every 10 min for 8-10 h.} During time lapse imaging, the cells were maintained at 37$^\circ$C and 5\% CO2 using a H301 stage top incubator with UNO controller (Okolab USA Inc, San Bruno, CA). After the imaging, cells were removed from the polyacrylamide substrates by incubating in 0.05\% trypsin for 20 min, and images of the fluorescent particles were collected; these images provided a traction-free reference state for computing cell-substrate tractions.

\subsection{Single cell tracking}
Images of GFP-labeled cell nuclei were segmented using StarDist, a plugin in ImageJ \cite{schmidt2018cell}. For computing cell velocities, we tracked the nucleus of each cell over time from time lapse imaging performed every 10 min. The centers of cell nuclei were identified from segmented images at each time point. For each time point, every cell nucleus was mapped to the nearest cell nucleus at the next time point using the knnsearch function in Matlab, allowing for cell trajectories and velocities to be computed. As the cell speed was typically no more than 0.5 $\mu$m/min and imaging was performed every 10 min, the average cell displacement was $< 5$ $\mu$m, which is much less than the average cell diameter of 22 $\mu$m. Hence, our method avoided mismatches between time frames. In the event of cell division, only one of the daughter cells was tracked. 
 
\subsection{Individual cell traction analysis}
Cell-induced displacements of the fluorescent particles were measured using Fast Iterative Digital Image Correlation \cite{bar2015fast} using $32\times32$ pixel subsets centered on a grid with a spacing of $8$ pixels ($2.6$ $\mu$m). Tractions were computed using unconstrained Fourier transform traction microscopy \cite{butler2002traction} accounting for the finite substrate thickness \cite{del2007spatio, trepat2009physical}. Next, a Voronoi tessellation was constructed using the center of each cell nucleus and the traction at each grid point was mapped to each cell. Finally, a vector sum of the traction within each cell was computed, giving the net traction applied by each cell. \textcolor{black}{For Supplemental Figs. S1d and S2b-c, the actual cell areas and outlines were determined by segmenting phase contrast images using SeedWaterSegmenter \cite{mashburn2012enabling}. From the segmented images, pixel coordinates of each cell were identified using the regionprops function in Matlab from which we computed the cell area and outlines.}

In the experiments that measured cell tractions and stained F-actin simultaneously, fluorescent particles in the substrate were imaged before seeding the cells, which provided the stress-free reference state for traction force microscopy. After seeding the cells and waiting 10-12 h for confluent islands to form, we imaged fluorescent particles in the deformed state of substrate with cells on it, then immediately fixed the cells, after which we stained for F-actin as described below.

\subsection{Monolayer stress microscopy}
Tensile stresses in the cell monolayer were computed using monolayer stress microscopy, which applies the principle of force equilibrium to the traction data to compute the two-dimensional stress tensor in the plane of the cell layer \cite{tambe2011collective, tambe2013monolayer}. Monolayer stress microscopy was performed as described in our recent manuscript \cite{saraswathibhatla2020spatiotemporal} using custom-written software that is freely available (https://github.com/jknotbohm/Cell-Traction-Stress).

\subsection{Chemical treatments}
Chemical treatments were cytochalasin D (Sigma-Aldrich), ML-141 (Sigma-Aldrich), Jasplakinolide (Sigma-Aldrich), and CN03 (Cytoskeleton, Inc, Denver, CO). Stock solutions of cytochalasin D, ML-141, and CN03 were prepared at 2 mM, 2 mM, and 0.1 g/L, respectively, all dissolved in dimethylsulfoxide except CN03 in water. The stock solutions were diluted in phosphate buffered saline to obtain desired concentrations for the experiments. 

\subsection{Stress fiber imaging and \textcolor{black}{orientation}}
For live imaging of stress fibers, the cells were seeded on a thin sheet of polyacrylamide ($< 20$ $\mu$m in thickness) and treated with 0.2 $\mu$M SiR-actin (Cytoskeleton, Inc). For fixed imaging of stress fibers, the cells were rinsed twice with PBS and fixed with 4\% paraformaldehyde in PBS for 20 min. The cells were washed with Tris-buffered saline twice for 5 min each and then incubated in 0.1\% Triton X-100 for 5 min at room temperature, and treated with 3-5 units/mL Phallodin Dylight 594 (Life Technologies catalog no.\ 21836). 

\textcolor{black}
{To compute the orientation of stress fibers in each cell, firstly, we computed the angle of the stress fiber image at each pixel using the OrientationJ plugin in ImageJ \cite{puspoki2016transforms}. Secondly, using the Voronoi-approximated cell outlines, the angles inside each cell were averaged to obtain the overall stress fiber orientation for each cell.}

\section{Traction alignment without alignment interaction}
Here we show that in the absence of alignment interactions there are no spatial correlations between tractions of neighboring cells.
In order to find the spatial and temporal correlation between traction vectors, we need to calculate $\left\langle \mathbf{n}\left(\mathbf{r},t\right)\cdot\mathbf{n}\left(\mathbf{r}^{\prime},t^{\prime}\right)\right\rangle$. We find,
\begin{eqnarray}
\left\langle \mathbf{n}\left(\mathbf{r},t\right)\cdot\mathbf{n}\left(\mathbf{r}^{\prime},t^{\prime}\right)\right\rangle  & = &\left\langle n_{x}\left(\mathbf{r},t\right)n_{x}\left(\mathbf{r}^{\prime},t^{\prime}\right)\right\rangle +\left\langle n_{y}\left(\mathbf{r},t\right)n_{y}\left(\mathbf{r}^{\prime},t^{\prime}\right)\right\rangle \nonumber \\
 & = &\left\langle \cos\theta\left(\mathbf{r},t\right)\cos\theta\left(\mathbf{r}^{\prime},t^{\prime}\right)\right\rangle +\left\langle \sin\theta\left(\mathbf{r},t\right)\sin\theta\left(\mathbf{r}^{\prime},t^{\prime}\right)\right\rangle \nonumber \\
 & = & \left\langle \cos\left(\theta\left(\mathbf{r},t\right)-\theta\left(\mathbf{r}^\prime,t^{\prime}\right)\right)\right\rangle,  \label{eq:n-n}
\end{eqnarray}
where in going from the second to the third line we used $\cos\left(\theta(\mathbf{r},t)-\theta(\mathbf{r}^\prime,t^\prime)\right)=\cos\left(\theta(\mathbf{r},t)\right)\cos\left(\theta(\mathbf{r}^\prime,t^\prime)\right)+\sin\left(\theta(\mathbf{r},t)\right)\sin\left(\theta(\mathbf{r}^\prime,t^\prime)\right)$. In the absence of alignment between neighboring cells, the first term on the right-hand-side in Eq. 2 vanishes (i.e., $\tau_a\to\infty$) and the equation of motion for the orientation of the traction vector $\mathbf{n}_i$ decouples from its neighbors. Therefore, for $\tau_a\to\infty$,
\begin{equation}
  \dot{\theta}_i = \eta_i.\label{eq:rot_no_align}
\end{equation}
In this case, it is straightforward to show that 
\begin{equation}
\left\langle \left(\theta_i\left(t\right)-\theta_j\left(t^{\prime}\right)\right)^{2}\right\rangle =\frac{2}{\tau_p}\left(t+t^{\prime}-2\psi\left(t,t^{\prime}\right)\right)\delta_{ij},\label{eq:theta2avg}
\end{equation}
where $\psi(x,y)$ is the minimum function that is equal to the smaller of the two of its arguments. 
We note that
\begin{equation}
t+t^{\prime}-2\psi\left(t,t^{\prime}\right)=\left|t-t^{\prime}\right|.
\end{equation}
If we recall that $\cos(\theta)=\mathrm{Re}\left(e^{i\theta}\right)$, and use the cumulant expansion, $\left\langle e^{i\theta}\right\rangle = e^{i\left\langle\theta\right\rangle -\frac{1}{2}\left(\left\langle\theta^{2}\right\rangle -\left\langle\theta\right\rangle ^{2}\right)}$, with $\langle\theta\rangle=0$, Eqs. (\ref{eq:n-n}) and (\ref{eq:theta2avg}) give
\begin{equation}
\left\langle \mathbf{n}\left(\mathbf{r},t\right)\cdot\mathbf{n}\left(\mathbf{r}^{\prime},t^{\prime}\right)\right\rangle = e^{-\left|t-t^{\prime}\right|/\tau_p}\delta\left(\mathbf{r}-\mathbf{r}^{\prime}\right).
\end{equation}
This shows that without alignment interaction, there is no spatial alignment between tractions of neighboring cells. In other words, any observed spatial correlation between alignments of neighboring cells would suggest presence of alignment interactions in the system.

\bibliographystyle{elsarticle-num} 
\bibliography{cell_migration}